\documentclass[fleqn,usenatbib]{mnras}

\usepackage{newtxtext,newtxmath}

\usepackage[T1]{fontenc}

\DeclareRobustCommand{\VAN}[3]{#2}
\let\VANthebibliography\thebibliography
\def\thebibliography{\DeclareRobustCommand{\VAN}[3]{##3}\VANthebibliography}

\usepackage{graphicx}
\usepackage{amsmath}	
\usepackage{xcolor}
\usepackage{tabularray}
\usepackage{booktabs}
\usepackage{tablefootnote}
\usepackage{threeparttable}
\usepackage{subfig}
\usepackage{array}
\usepackage{tabularx}

\title[On the spin evolution of older Sun-like stars]
{Why the observed spin evolution of older-than-solar like stars might not require a dynamo mode change}

\author[K. Kotorashvili et al.]{Ketevan Kotorashvili $^{1,2}$\thanks{kkotoras@ur.rochester.edu}, Eric G.~Blackman$^{1,2}$\thanks{eric.blackman@rochester.edu},{James E.~Owen}$^{3}$\thanks{james.owen@imperial.ac.uk}\\
$^{1}$Department of Physics and Astronomy, University of Rochester, Rochester NY 14627\\
$^{2}$Laboratory for Laser Energetics, University of Rochester, Rochester, NY 14623, USA \\
$^{3}$Astrophysics Group, Department of Physics, Imperial College London, Prince Consort Rd, London SW7 2AZ, UK\\}
\pubyear{2022}
\begin{document}
\label{firstpage}
\pagerange{\pageref{firstpage}--\pageref{lastpage}}
\maketitle

\begin{abstract}
The spin evolution of main sequence stars has long been of interest for basic stellar evolution, stellar aging, stellar activity, and consequent influence on companion planets.  Observations of older than solar late-type main-sequence stars have been interpreted to imply that a change from a dipole-dominated magnetic field to one with more prominent higher multipoles  might be necessary to account for the data.   The spin-down models that lead to this inference are essentially tuned to the Sun.  
Here we take a different approach which considers 
individual stars as fixed  points rather than just the Sun. We  use a time-dependent theoretical model to solve for the spin evolution of low-mass main-sequence stars that includes a Parker-type wind and a time-evolving magnetic field coupled to the spin.
Because the wind is exponentially sensitive to the stellar mass over radius and the coronal base temperature,  the use of each observed star  as a separate fixed point is more appropriate  and, in turn, produces a set of  solution curves that produces a solution  envelope rather than a simple line.  This envelope of solution curves, unlike a single line fit, is consistent with the data and
does not unambiguously require a modal transition in the magnetic field to explain it.

\end{abstract}


\section{Introduction} \label{sec:intro}
Understanding the coupled spin-activity evolution of stars is of interest both for the basic physics of  rotating stellar evolution and stellar activity, for determining stellar ages via gyrochronology, and for quantifying the influence of stellar activity on companion planetary atmospheres.   Predicting the spin evolution of main sequence stars and the associated activity ultimately requires an accurate model for the coupled evolution of their magnetic fields, their spin, their activity and mass loss.

Until recently the standard period-age evolution for main sequence solar-like FGK stars has been divided into two regimes, saturated and unsaturated. The empirically determined  transition between them occurs at  $\Tilde{R}o \sim 0.13$, where the Rossby number $\Tilde{R}o$  is defined as $\Tilde{R}o = P/\tau_c$,  with $P$ being the star's rotation period and $\tau_c$ the stellar  model-inferred convective turnover time \citep{Wright2011,Reiners2014}. Very young, X-ray luminous  stars are in the saturated regime where their X-ray to bolometric luminosity ratio is nearly  independent of rotation rate.  Older stars are in the unsaturated regime for which the period age relation has  been traditionally  characterized by the empirical Skumanich law \citep{Skumanich1972}.
Recently however, for a sub-population of stars older than the Sun,  the  spin-down rate has been purported to be slower than that of the Skumanich law \citep{Skumanich1972} and slower than that predicted by some standard spin-down models with a fixed magnetic field geometry \citep{Matt2012,Reiners2012Mohanty,vanSaders2013,Gallet2013improved,Matt2015,vanSaders2016}.  This has led to the suggestion that  dynamos in these stars may be incurring a state transition from  dipole  to one in which the field is dominated by  higher multipoles or otherwise weaker field that less effectively removes angular momentum \citep{vanSaders2016,Tripathi+2021}.  Such a transition would then warrant a theoretical explanation. 

The importance of this potential transition  warrants further investigation to assess whether  it is unambiguous.   In particular, how precise are the predictions of spin evolution from current theoretical models that invoke no dynamo transition, and how are these models used to obtain a predicted envelope of 
spin-period evolution bounds for the evolution of a population of stars similar to, but not identical to, the Sun? 

To address this,  we study the time evolution of the rotation period for older-than-solar late-type stars
using an example theoretical model for the coupled time evolution of the X-ray luminosity, magnetic field strength, mass loss and rotation. Importantly, the  observed data for each star provides  boundary conditions needed to  solve the system of equations for each specific star.  We do not assume that each star is an identical twin to the Sun. This distinction proves to be important in limiting the precision of what can be inferred   and  the robustness of whether the observations definitively reveal the need for a dynamo transition in each star. 


In Section \ref{sec:model equations}, we summarize the minimalist theoretical model that couples the time evolution of X-ray luminosity, rotation, magnetic field and mass loss \citep{Blackman2016}. In Subsection \ref{sec:xray and mass loss} we provide expressions for X-ray luminosity and mass loss as a function of the X-ray coronal temperature for cases when thermal conduction is dominant and when thermal conduction can be ignored. Thermal conduction can reduce the hot gas supply to the wind, lowering its ability to spin down the star, but also keeps the magnetic field stronger longer which would exacerbate spin down. The net effect of this competition has yet to be quantified.  In Section \ref{sec:solutions} we obtain solutions  for the time evolution of the rotation period of each individual star in a sample of old stars with observed spins and ages, using their observed stellar properties as fixed point boundary conditions for the solutions. 
We find that even the small variations in  observed properties (e.g. magnetic field, mass, radius) between solar-like stars, makes fitting an evolution model to a single star like the Sun  not sufficiently representative of the population to identify that the population as a whole is incurring a dynamo transition.
We conclude in Section \ref{sec:conclusion} and address some  broader implications for comparing  theory and observation.

\section{Physical Model and Equations}\label{sec:model equations}

Main sequence low-mass stars spin down as a consequence of their magnetized stellar winds \citep{Parker1958,Schatzman1962,Weber1967,Mestel1968}. F, G , K and M stars with  masses in the range $0.35M_{\odot} < M < 1.5M_{\odot}$  have a  radiative core surrounded by a convective envelope and are in that respect potentially most solar-like with respect to their dynamos \citep{Parker1955,Steenbeck1969}. 
The magnetic field anchors 
the stellar wind to the surface of the star, forcing it to 
co-rotate  up to   the Alfv\'en radius,
so angular momentum is lost from the star. As a result, the reduced angular momentum means reduced free energy available for the dynamo, and the magnetic field and X-ray luminosity also decrease.
Therefore the strength of the magnetic field at the surface, the rate of angular momentum loss, X-ray luminosity and the rotation period are fundamentally linked \citep{Kawaler1988}.

Here we use and adapt a minimalist holistic model for this coupled  time evolution of X-ray luminosity, mass loss, rotation and magnetic field strength \citep{Blackman2016} to explain the ﬂattening in the observed period–age relation for older stars than the Sun. In this model, some fraction of dynamo-generated magnetic field lines are considered open, allowing stellar wind to remove angular momentum, while some fraction of field lines are considered closed, sourcing the thermal X-ray emission.
 The magnetic field expression is based on a dynamo saturation model in a regime 
 where the total saturated field strength depends on the rotation rate 
 The  dynamo-produced magnetic field is then mutually evolving  with the spin evolution of low-mass main-sequence stars in this slow rotator regime.  

In this section, we briefly summarize 
the minimalist theoretical model that couples the time evolution of the aforementioned stellar properties, discuss the main ingredients of the model, and point out a few numerical coefficient corrections to previous work.
We also apply the formalism for stars other than the Sun  and use the properties of each individual star for which 
we have observed data as a boundary condition for respective solutions.  The importance of this as it pertains to making the theoretical prediction of spin-down with age  an "envelope" rather than a "single line" will be exemplified and emphasized later in the paper.
We  provide only the streamlined set of resulting equations here, and the detailed derivations of the original model equations on which our revised derivations are based can be found in \cite{Blackman2016}.



\subsection{Saturated magnetic field and X-ray luminosity}

 The dynamo-produced magnetic fields are estimated \citep{Blackman2015, Blackman2016} by:  (1) using a generalized correlation time for dynamos that equals the convection time ($\tau_{c}$) for slow rotators and becomes proportional to the rotation time
 for fast rotators
 and (2) using a  dynamo saturation model,  based on the combination of magnetic helicity evolution and loss of magnetic field by magnetic buoyancy \citep{Blackman2002,Blackman2003}. In the slow-rotator regime of interest, the field saturation depends on the rotation rate, but the exact  
field saturation model is less important than the fact  that there remains a spin dependence of the field strength and that the saturation time (of order cycle period) is short compared to the Gyr time scales of secular evolution we are interested in.
This results in the expression for normalized surface radial magnetic field:

\begin{equation}
    b_r(t) \equiv \frac{B_{r*}(t)}{B_{r,*n}}=g_L(t)  \sqrt{ \frac{1+s\Tilde{R}o_*}{1+s\Tilde{R}o(t)}},
	\label{eq:br}
\end{equation}
 where $B_{r,*n}$ is present-day radial magnetic field value for each star (here $n$ indicates  "now")  and $\Tilde{R}o_*$ represents the present-day Rossby number for each star.
 The factor $g_L(t)=\left(\frac{1}{1.4-0.4t}\right)^{\frac{\lambda-1}{4}}$ approximates the fusion-driven increase in the bolometric luminosity $\mathcal{L}_{bol}$with time $t$ in units of solar age from solar models \citep[e.g.][]{Gough1981}, and deviates from unity only if $\mathcal{L}_{bol}$ evolves. We  crudely apply the same approximation for $g_L(t)$ to other solar-like stars scaled in terms of their age.\footnote{More detailed  empirical fits for each stellar model could be inferred but this is beyond the level of precision required for present purposes.} Here $s$ is a shear parameter defined as $|\Omegaup_0-\Omegaup(r_c,\theta_s)|=\Omegaup_0/s$, where $\Omegaup$ is surface rotational speed; $\theta_s$ is a fiducial polar angle; $r_c$ is a fiducial radius in the convective zone and $ \lambda$ is a parameter  representing the power law dependence of the magnetic starspot area covering fraction $\Theta$  on X-ray luminosity $\mathcal{L}_X$, namely
 $\Theta\propto \mathcal{L}_X^\lambda$.
 In our case, we take $\lambda=1/3$,  consistent with the range inferred from observations of star spot covering fractions \citep{NicholsFleming+2020} and we fix the shear parameter at $s=8.3$, because the transition from the saturated to the unsaturated regime of X-ray luminosity was best matched theoretically with this value  \citep{Blackman2015,Blackman2016}. In practice, this has to be determined with detailed calculations, but the specific value does not affect the overall message of the present paper as our focus is  on the unsaturated regime where the shear term contribution to the correlation time is small.

The estimated X-ray luminosity derived in \cite{Blackman2015} is the product of the magnetic energy ﬂux, averaged over the  change over a stellar cycle for Sun-like stars \citep{peres2000}, times the surface area through which the magnetic field
penetrates the photosphere. The result from that calculation is
\begin{equation}
    \mathcal{L}_X = \mathcal{K} \mathcal{L}_{mag} \simeq \mathcal{K} \frac{2}{3} \bigg(\frac{B^2_{\phi}}{8 \pi}\bigg)^2 \frac{\Thetaup r^2_c}{\rho \varv},
	\label{eq:lxmag}
\end{equation}
 where $\rho$ and $\varv$ are a typical density and turbulent velocity in the convection zone; and $\mathcal{K}$  is the fraction of the area-integrated magnetic energy flux    $\mathcal{L}_{mag}$, that goes to into X-ray luminosity. The quantity $\mathcal{L}_{mag}$ is 
estimated in equation (\ref{eq:lxmag}). 
 In \citep{Blackman2016} $\mathcal{K}$ was approximated  as $1/2$ based on the coronal equilibrium solution when conduction is unimportant. We find this is also an acceptable  approximation when conduction dominates so we adopt it. This 
 leads to the  relation between X-ray luminosity and radial magnetic field  (\cite{Blackman2016}):
\begin{equation}
    l_x \equiv \frac{1}{1.4-0.4t}\bigg(\frac{1+s\Tilde{R}o_*}{1+s\Tilde{R}o(t)}\bigg)^{\frac{2}{1-\lambda}}=b^{^{\frac{4}{1-\lambda}}}_r.
	\label{eq:lxbr}
\end{equation}



\subsection{Angular velocity evolution}

 \cite{Blackman2016}  considered  angular momentum loss by the stellar wind in the equatorial plane and used the  (\cite{Weber1967}) model to find the surface toroidal magnetic field
 and  the equation for angular velocity. Following derivations in \cite{Weber1967}, \cite{Lamers1999} and \cite{Blackman2016} for the Alfv\'en radius  $r_A$ we have
 \begin{equation}
 \frac{r_A}{R_*}=\left(1-\frac{R_*B_{r*}B_{\phi*}}{\Dot{M}\Omega_*}\right)^{1/2}=\left(1+\frac{R_*|B_{r*}||B_{\phi*}|}{\Dot{M}\Omega_*}\right)^{1/2},
\label{eq:ra1}
\end{equation}
where  $R_*$ is a stellar radius and 
$\Dot{M}= 4\pi \rho r^2 U_r$, where we used Parker wind solutions \citep{Parker1955} for a radial wind speed $U_r$ . 
Compared to the same equation in \cite{Blackman2016},  we emphasize that there is a positive sign when absolute values are used because of  the opposite signs of $B_{\phi*}$ and  $B_{r*}$.

A separate equation for $ \frac{r_A}{R_*}$ derived  from the mass loss rate to outflow speed relation, the definition of the Alfv\'en radius, and the radial field fall off of $1/r^2$ is,
\begin{equation}
 \frac{r_A}{R_*}=\frac{b_{r*}}{\Dot{m}^{1/2}\Tilde{u}^{1/2}_A}\frac{R_*B_{r,*n}}{\Dot{M}^{1/2}_{*n} u^{1/2}_{A,*n}},
\label{eq:ra}
\end{equation}
which when combined with equation (\ref{eq:ra1}), gives
\begin{equation}
    b_{\phi*}\equiv\frac{B_{\phi*}(t)}{B_{\phi ,*n}}=-\frac{\Dot{m}\omega_*}{b_{r_*}} \frac{M_{*n}\Omega_{*n}}{R_*B_{\phi,*n}B_{r,*n}}\left[\frac{r^2_A}{R^2_*}-1\right],
	\label{eq:bfi}
\end{equation}
where $ \Dot{M}_{*n}$ and $B_{\phi,*n}$ are the present-day mass loss and  toroidal magnetic field values for each star; $\Dot{m}$ is a mass loss  derived later (see equations (\ref{eq:lxm1}) and  (\ref{eq:lxm2}) for regime I and regime II respectively); $ M_{*n}$ is the present-day individual stellar mass; and $\omega_{*}(t)=\Omega(t)/\Omega_{*n}$, where $\Omega_{*n}$ represents the present day value of angular velocity for each individual star. For the Sun, $\Omega_{*n}=\Omega_{\odot}=2.97 \cdot 10^{-6} /s$,
 $B_{\phi,*n}=B_{\phi\odot}=1.56 \cdot 10^{-2} G$, $B_{r,*n}=B_{r\odot}=2 G$. For other stars, the corresponding values in Table \ref{table:datapoints} will be used.
In equation (\ref{eq:ra}), $\Tilde{u}_A(t)$ is the normalized Alfv\'en speed given by  
\begin{equation}
    \Tilde{u}_A(t) \equiv \frac{u_A}{u_{A,*n}}= \sqrt{\frac{T_*}{T_{*n}}}
    \frac{W_k[-D(r_A)]}{W_k[-D(r_{A,*n})]},
	\label{eq:ua}
\end{equation}
where $T_*$ is the coronal X-ray temperature and $T_{*n} $ is the coronal X-ray temperature at present time (now) for each specific star. $W_k[-D(r_A)]$ is the Lambert W function for Parker wind solutions $k= 0$ for $r\leq r_s$ and $k= -1$ for $r\geq r_s$ \citep{Cranmer2004} and 
\begin{equation}
   D(r_A)=\left(\frac{r_A}{r_s}\right)^{-4} {\rm exp} \left[4\left(1-\frac{r_s}{r_A}\right)-1\right].
\end{equation}
The sonic radius is given by
\begin{equation}
 \frac{r_s}{R_*}=\frac{GM}{2c^2_sR_*}   
\end{equation}
with isothermal sound speed $c_s \propto T^{1/2}$.

The evolution of stellar angular velocity in  dimensionless form is given by 
\begin{equation}
    \frac{d\omega_*}{d\tau} \equiv - \omega_* \frac{q}{0.059}\frac{b^2_r}{ m_* \Tilde{u}_A}\frac{B^2_{r,*n}R^2_*\tau_{*n}}{M_{*n}u_{A,*n}},
	\label{eq:omegadot}
\end{equation}
where $\tau_{*n}$ is present-day age for each individual star; $q$ is the inertial parameter, that depends on internal angular momentum transport and defines what fraction of the star contributes to the spin-down (and corrected a 
 typo on the right of equation (41) of  \cite{Blackman2016} which had residual factor of $\Omega_\odot$  and was missing the $R^2_*/0.059$ factor.).
We use $q=1$ for all stars, which  indicates a conventional assumption that the field is  coupled to the moment of inertia of the  full stellar mass. This could in principle be violated if the field were not anchored sufficiently deeply and angular momentum transport within the star was inefficient.

\subsection [\texorpdfstring{$\mathcal{L}_x, \Dot{M} $ and $T_0$}{(Lx, Mdot,t0)}]	{Coronal Equilibrium: relation between $\mathcal{L}_x, \Dot{M}$ and $T_0$}\label{sec:xray and mass loss}

The above equations show  that X-ray luminosity, dynamo-produced magnetic field  and angular velocity are all coupled. To determine how all of these 
quantities are connected to the mass loss rate, we follow the procedure of \cite{Blackman2016} but since that paper focused on younger-than-solar stars,  here we study both younger and older stars and generalize the equations accordingly. 


 Magnetic fields are the source of input energy to the corona in our model, which is then distributed into either winds, x-rays, or lost to the photosphere by thermal conduction.  
Equilibrium is established between the sinks of mass loss, X-ray radiation and conduction over time scales short compared to spin-down time scales and  can be used to determine the dominant sinks of the magnetic energy flux. 
 
 According to \cite{Hearn1975}, for a given coronal base pressure, there is an average coronal temperature that minimizes energy loss. The minimum coronal flux  condition is given by
\begin{equation}
    \frac{\partial}{\partial T}(F_{W1}+F_c+F_x)=\frac{\partial}{\partial T}F_B=0,
	\label{eq:equilibrium}
\end{equation}
where $F_B$ is the  flux of magnetic energy sourced into the coronal base and 
$F_{W1}, F_c, F_x$ are respectively  the wind flux, conductive loss, and the radiative (X-ray) loss, from the one density scale height region above the chromosphere.

The expression for coronal energy loss in the stellar wind is given by
\begin{equation}
    F_{W1}=3.1 \times 10^6 p_0 \Tilde{T}^{1/2}_* e^{3.9\frac{m_*}{r_*}\left(1-\frac{1}{\Tilde{T_*}}\right)} {\frac{\rm erg}{\rm cm^2\cdot s}},
	\label{eq:fw}
\end{equation}
where we used the isothermal Parker wind solution \citep{Parker1955}   along with the assumption of large-scale magnetic fields being approximately radial  out to the Alfv\'en radius ($r_A$). Here  $\Tilde{T}_{*}=\frac{T_*}{T'_*}$ is a dimensionless temperature with a different normalization parameter $T'_*$ for each star;  $m_*=\frac{M}{M_{*n}}$
and  $r_*=\frac{R_0}{R_{*n}}$, where $R_0$ and $ R_{*n}$ represent radius at the coronal base and a specific individual stellar radius.
Normalizing  stellar parameters to individual stars, we then have $m_*=r_*=1$. We also use $p_0\sim\rho_0c^2_s$ where the subscript 0 indicates values at the coronal base and we use  CGS units for  $p_0$.


For the X-ray radiation flux, we have
\begin{equation}
    F_x=1.24 \times 10^6 \frac{p_0^2}{\Tilde{T}^{5/3}_*}\frac{r^2_*}{m_*} {\frac{\rm erg}{\rm cm^2\cdot s}},
	\label{eq:fx}
\end{equation}

For the conductive loss, 
\begin{equation}
    F_c=4.26 \times 10^6 p_0\Tilde{T}^{3/4}_*\frac{\Tilde{{\Thetaup}}}{4\pi}{\frac{\rm erg}{cm^2\cdot s}},
	\label{eq:fc}
\end{equation}
where the solid angle correction fraction $\frac{\Tilde{{\Thetaup}}}{4\pi} \leq 1$ arises because conduction down from the corona is assumed to be  non-negligible only along the fraction of the solid angle covered with field lines perpendicular to the surface. 

There is a monotonic relation between the base pressure of the corona and the energy
density at coronal equilibrium, and all three energy losses increase with the base coronal pressure. The above equations lead to an equilibrium pressure (with corrected numerical coefficients in the first and third term, as well as the corrected factor of $\frac{m_*}{r_*}$ in the last term compared to \cite{Blackman2016})
\begin{equation}
\begin{split}
    p_0=\frac{m_*}{r^2_*}0.12 \Tilde{{\Thetaup}}\Tilde{T}_{0*}(t)^{\frac{29}{12}}&+\frac{m_*}{r^2_*}0.75\Tilde{T}_{0*}(t)^{\frac{13}{6}}e^{3.9\frac{m_*}{r_*}\left(1-\frac{1}{\Tilde{T}_{0*}(t)}\right)} \\ &+\frac{m^2_*}{r^3_*}5.85\Tilde{T}_{0*}(t)^{\frac{7}{6}}e^{3.9\frac{m_*}{r_*}\left(1-\frac{1}{\Tilde{T}_{0*}(t)}\right)},
	\label{eq:p0}
\end{split}
\end{equation}
where $\Tilde{T}_{0*}(t)={\frac{T_{0*}(t)}{T'_*}}$, $T_{0*}(t)$ is the  temperature at equilibrium for each specific star. It is derived from   equation (\ref{eq:equilibrium}) and equation (\ref{eq:p0}). We assume that this equilibrium is established on a time scale that is short compared to the secular gigayear evolution times of interest. For the  present solar coronal temperature we take $T_{0,*n} \sim {T}_{\odot} \sim 1.5 \times 10^6 K$ and for $T'_* $ we used $T'_* =T'_\odot = 3 \times 10^6 K$, so that at $ \Tilde{T}_{0*}=0.5, \ l_x=\frac{\mathcal{L}_x}{\mathcal{L}_{x,*n}}=1$ and $\Dot{m}=\frac{\Dot{M}}{\Dot{M}_{*n}}=1$.

\begin{figure}
	\includegraphics[width=\columnwidth]{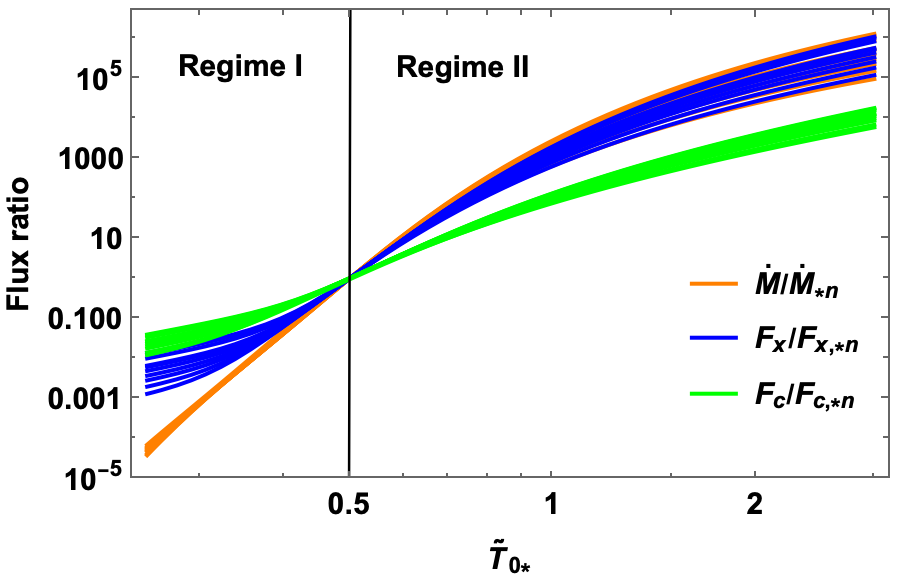}
    \caption{Normalized energy fluxes of X-rays $\frac{F_x}{F_{x,*n}}$ (blue); thermal conduction $\frac{F_c}{F_{c,*n}}$ (green);, and mass outflow   $\frac{\Dot{M}}{{M}_{*n}}$ (orange) are shown for each individual star of Table \ref{table:datapoints}.  Similar plots  were shown  in \protect\cite{Blackman2016} but only for the Sun. The y-axis is in units of individual stellar values for each quantity and the unobserved equilibrium temperature ${T}_{0*}$ for each star is normalized such that a transition between the dominance and sub-dominance  of thermal conduction occurs at dimensionless $\Tilde{T}_{0*}=0.5$.  In regime I, ($\Tilde{T}_{0*} < 0.5$), thermal conduction is dominant, but it is subdominant in regime II ($\Tilde{T}_{0*} > 0.5$), where $l_x \simeq \Dot{m}$. Regime I corresponds to older and regime II to younger phases of the main sequence for a given star. The envelope of these curves for the different stars produces the  bands of color for each  energy flux.}
    \label{fig:hearn}
\end{figure}

Fig.\ref{fig:hearn} shows radiation, conduction and total coronal wind fluxes 
 $\frac{F_x}{F_{x,*n}}$, $\frac{F_c}{F_{c,*n}}$, $\Dot{m}=\frac{F_{W1}\Tilde{T}_{0,*n}}{F_{W1,*n}\Tilde{T}_{0*}}$ as function of equilibrium temperature, where $\Tilde{T}_{0,*n}$ is the coronal temperature at present time for Sun-like stars.  All the quantities (y-axis) and the equilibrium temperature (x-axis) are normalized to their respective stellar values for an individual star.   We define Regime I as the lifetime phase of a star for which thermal conduction flux dominates the outflow flux and Regime II when the reverse is true. This occurs at a different coronal equilibrium temperature $T_{0*}$ specific to each star.  We then define the  transition to occur at the same arbitrary dimensionless value of $0.5$ for each star such that  
 $\Tilde{T}_{0*}<0.5$ corresponds to regime I and $\Tilde{T}_{0*}>0.5$ corresponds to regime II. 
 The vertical line at $\Tilde{T}_{0*}=0.5$ represents the transition between the two regimes 
 which have different relations between X-ray luminosity and mass loss.

\subsubsection{Regime I (conduction dominated)}

In this regime, which generally corresponds to the spun-down older main-sequence phase of a given star, the  first term of equation \eqref{eq:p0} dominates. Consequently, the normalized value for the X-ray luminosity is $l_x=\frac{\mathcal{L}_x}{\mathcal{L}_{x,*n}}=\frac{F_x}{F_{x,*n}}$, which, for each star can be written
\begin{equation}
l_x \simeq \left(\frac{\Tilde{T}_{0*}}{{\Tilde{T}_{0,*n}}}\right)^{\frac{19}{6}}. 
\end{equation}
The normalized mass loss is $\Dot{m}=\frac{\Dot{M}}{\Dot{M}_{*n}}$
\begin{equation}
\Dot{m} \simeq \left(\frac{\Tilde{T}_{0*}}{{\Tilde{T}_{0,*n}}}\right)^{\frac{23}{12}} 
\exp \bigg[\frac{3.9}{\Tilde{T}_{0*}}\frac{m_*}{r_*}\bigg(\frac{\Tilde{T}_{0*}}{\Tilde{T}_{0,*n}}-1\bigg)\bigg],
\label{eq:lxm1}
\end{equation}
which  couples  with the three other stellar properties discussed above.

\subsubsection{Regime II (no conduction)}

In this regime, which generally corresponds to the younger, faster-rotating phase of a given star, the  second term on the right of equation \eqref{eq:p0} dominates, which is the outflow flux term. So for $l_x$ and $\Dot{m}$ we have  \citep{Blackman2016}
\begin{equation}
    l_x \simeq \exp \bigg[\ln(\Tilde{T}_{0*}) +\frac{7.8}{\Tilde{T}_{0*}}\frac{m_*}{r_*}\bigg(\frac{\Tilde{T}_{0*}}{\Tilde{T}_{0,*n}}-1\bigg) \bigg] \simeq \Dot{m}.
	\label{eq:lxm2}
\end{equation}



\section{Time-evolution of rotation period}\label{sec:solutions}

We numerically solved the four equations  (\ref{eq:lxbr}), (\ref{eq:bfi}), (\ref{eq:omegadot}) and (\ref{eq:lxm1}) or (\ref{eq:lxm2}) respectively for regimes I and II, along with equations (\ref{eq:ra}) and (\ref{eq:ua}) for the spin evolution.
Importantly, we solved these equations for individual stars, using measured stellar properties as a fixed point (boundary condition) corresponding to the observations of that particular star. The set of  solutions  comprises an envelope of these individual curves.

\subsection{Solutions and comparison to data}

\defcitealias{smith2014}{Paper~I}
\begin{table*}
\begin{threeparttable}[b]
\caption{Stellar properties of G-type and F-type stars used in our study (\protect\cite{wright2004chromospheric,Bazot2012bayesian,Zakowicz2013spectral,Marsden2014,vanSaders2016,Creevey2017data,white2017,Metcalfe2022})}
\begin{tabular}{llllllllll} 
    \toprule
     & {KIC ID/Name} & {Sp.} &{Radius} & {Mass} & {Age} & {Period} & {Luminosity} & {Rossby} & {Magnetic field} \\
     & {or HIP no.} & {Type} &{$(R_{\sun})$} & $(M_{\sun})$ & {(Gyr)} & {(Days)} & {$(L_{\sun})$} &  {number} & {(G)} \\
     \midrule
    1  & Sun & {G2V} & {1.001 ± 0.005} & {1.001 ± 0.019} & {4.6} & {24.47} & {0.97 ± 0.03} & {2} & {2}\\
    2  & {9098294}  & {G3V} & {1.150 ± 0.003} & {0.979 ± 0.017}  & {8.23 ± 0.53} & {19.79±1.33} & {1.34 ± 0.05}  & {\tnote{*}} & {\tnote{*}} \\
    3  & {7680114}  & {G0V } & {1.402 ± 0.014} & {1.092 ± 0.030}  & {6.89 ± 0.46} & {26.31±1.86} & {2.07 ± 0.09}   & {\tnote{*}} & {\tnote{*}}  \\
    4  & {$\alpha$ Cen A}   & {G2V} & {1.224 ± 0.009} & {1.105 ± 0.007}   & {5.40 ± 0.30} & {22±5.9} & {1.55 ± 0.03} & {\tnote{*}} & {\tnote{*}}\\
    5  & {16 Cyg-A\tnote{$a$}}   & {G1.5Vb} & {1.223 ± 0.005} & {1.072 ± 0.013}  & {7.36 ± 0.31} & 
    $20.5${\raisebox{0.5ex}{\tiny$^{+2}_{-1.1}$}} & {1.52 ± 0.05}   & {\tnote{*}} & {< 0.5} \\
     6  & {16 Cyg-B\tnote{$a$}}   & {G3V} & {1.113 ± 0.016} & {1.038 ± 0.047}  & {7.05 ± 0.63} & 
    $21.2${\raisebox{0.5ex}{\tiny$^{+1.8}_{-1.5}$}} & {1.21 ± 0.11}   & {\tnote{*}} & {< 0.9} \\
    7  & {18 Sco\tnote{$a$}}   & {G2Va} & {1.010 ± 0.009} & {1.020 ± 0.003}  & 
    $3.66${\raisebox{0.5ex}{\tiny$^{+0.44}_{-0.5}$}}& {22.7 ± 0.5} & {1.07 ± 0.03 }   & {\tnote{*}} & {1.34} \\
     \midrule
    8  & {1499}  & {G0V} & {1.11 ± 0.04} & {$1.026${\raisebox{0.5ex}{\tiny$^{+0.04}_{-0.03}$}}} & {$7.12${\raisebox{0.5ex}{\tiny$^{+1.40}_{-1.56}$}}}
   & {$29$}{\raisebox{0.5ex}{\tiny$^{+0.3}_{-0.3}$}}  & {1.197} & {2.16} & {0.6 ± 0.5}\\ 
    9 & {682}  & {G2V} & {1.12 ± 0.05} & {1.045}{\raisebox{0.5ex}{\tiny$^{+0.028}_{-0.024}$}}  & 
     {$6.12$}{\raisebox{0.5ex}{\tiny$^{+1.28}_{-1.48}$}}& {$4.3$}{\raisebox{0.5ex}{\tiny$^{+0.0}_{-0.2}$}} & {1.208} & {0.4} & {4.4 ± 1.8}\\
    10  & {1813}  & {F8} & {1.18}{\raisebox{0.5ex}{\tiny$^{+0.06}_{-0.05}$}} & {0.965}{\raisebox{0.5ex}{\tiny$^{+0.02}_{-0.02}$}}  & 
   {10.88}{\raisebox{0.5ex}{\tiny$^{+1.36}_{-1.36}$}}  & {$22.1$}{\raisebox{0.5ex}{\tiny$^{+0.2}_{-0.2}$}} & {1.315}  & {1.95} & {2.4 ± 0.7} \\
    \midrule
     11  & {176465 A }   & {G4V} & {0.918 ± 0.015} & {0.930 ± 0.04}  & {3.0 ± 0.4} & {19.2±0.8 } & {}   & {}  & {}  \\  
    12 & {176465 B }   & {G4V} & {0.885 ± 0.006} & {0.930 ± 0.02}  & {2.9 ± 0.5} & {17.6±2.3} & { }   & {}  & {}     \\
     13  & {400}  & {G9V} & {0.8}{\raisebox{0.5ex}{\tiny$^{+0.02}_{-0.03}$}} & {0.794}{\raisebox{0.5ex}{\tiny$^{+0.034}_{-0.018}$}}  & {12.28{\raisebox{0.5ex}{\tiny$^{+1.72}_{-7.08}$}}} & {$35.3$}{\raisebox{0.5ex}{\tiny$^{+1.1}_{-0.7}$}} & {0.455} & {2} & {2.1 ± 1.0}\\
    \midrule
    14 & {6116048}   & {F9IV-V} & {1.233 ± 0.011} & {1.048 ± 0.028}  & {6.08 ± 0.40} & {17.26±1.96} & {1.77 ± 0.13 }   & {}  & {}     \\
    15  & {3656476}  & {G5IV} & {1.322 ± 0.007} & {1.101 ± 0.025}  & {8.88 ± 0.41} & {31.67±3.53} & {1.63 ± 0.06}  &  
    \label{table:datapoints}\\
    \toprule
\end{tabular}
 \begin{tablenotes}
       \item [$a$] For 16 Cyg-A, 16 Cyg-B and 18 Sco we used estimated mass loss rates from \cite{Metcalfe2022}, based on the scaling relation $\Dot{M} \simeq F_x^{0.77±0.04} $ \citep{Wood2021}. For other stars we have used the Solar value.  
       \item [*] In our solutions we have used Solar values for these parameters.
     \end{tablenotes}
     \end{threeparttable}
\end{table*}

\begin{figure}
\centering
\subfloat{\includegraphics[width=\columnwidth]{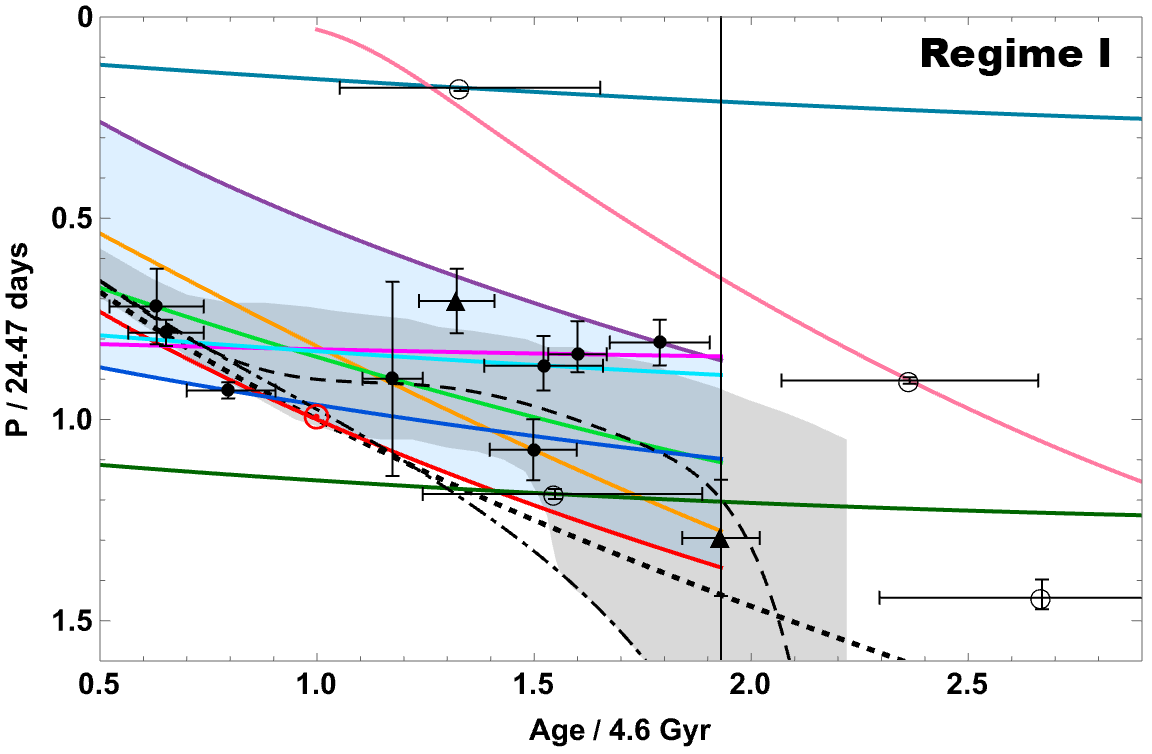}}  \\
\subfloat{\includegraphics[width=\columnwidth]{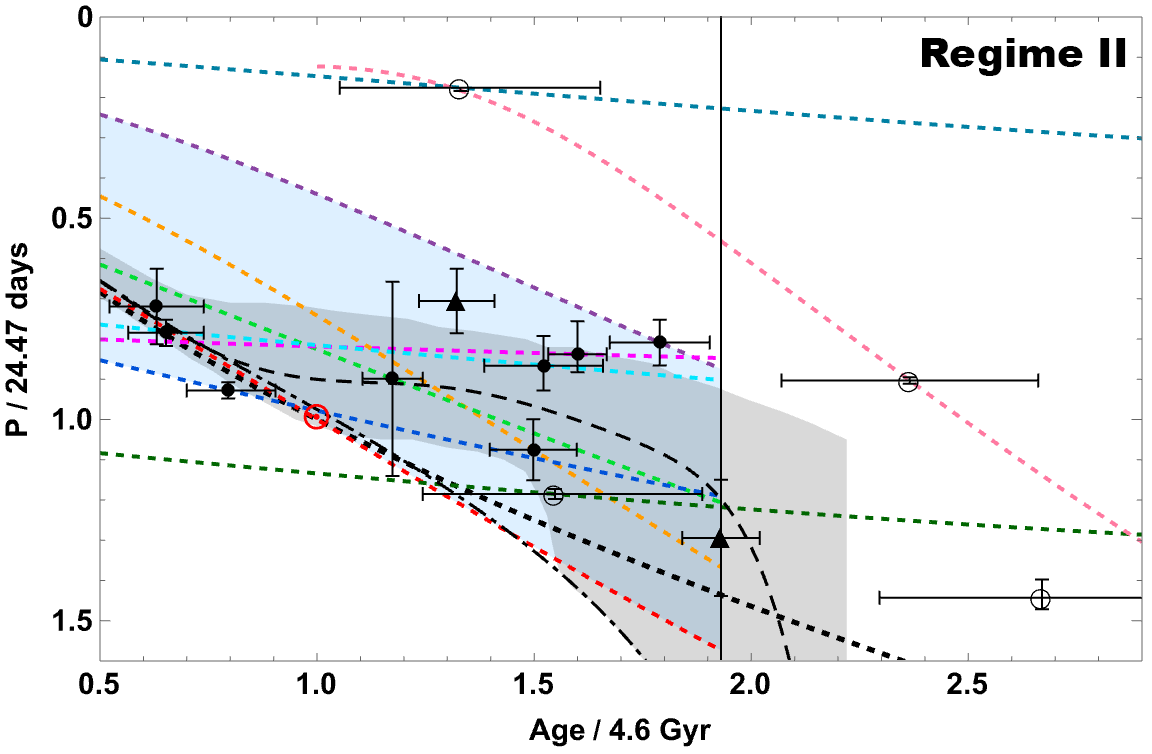}}
 \caption{The two panels show envelopes of solution curves for the time evolution of the rotation period,  where each observed star is a fixed point on the individual curve. Panel a and b correspond to the regime I and regime II solutions respectively, where the y and x-axis are normalized to solar period and age. Data points and boundary conditions used to find individual solution curves are given in Table \ref{table:datapoints}. Corresponding solutions for row numbers therein are color-coded as 1 - red,  2 - purple, 3 - orange, 4 - green, 5 - magenta, 6 - cyan, 7 - blue,  8 - dark green, 9 - dark cyan and 10 - pink. Open circles correspond to data points from the Bcool project magnetic survey (respectively 8, 9, 10 and 13 from Table \ref{table:datapoints}). \citep{Marsden2014}. The Sun is marked as red $\odot$. Triangles represent a star transitioning from the main-sequence to the subgiant phase and a subgiant (respectively 14 and 15 from Table \ref{table:datapoints}). The vertical line represents the cutoff before the subgiant phase for the stars 1-7 in  Table \ref{table:datapoints}. The blue-shaded region represents the envelope of solutions for all the stars except the ones with large uncertainties in age from the Bcool project. Both regime I and regime II solutions are compared with the Skumanich law (black dotted line), a standard rotational evolution model (black dot-dashed line) \citep{vanSaders2016}, a modified rotational evolution model (black dashed line) and the gray shaded region  \citep{metcalfe2017} that represents the expected dispersion due to different masses, metallicities and effective  temperatures between 5600-5900 K.
}
 \label{fig:solution}
\end{figure}

Data Table \ref{table:datapoints} shows the properties of the  G-type and F-type stars available for the study. Most of the G stars come from a  sample from 21 Kepler with  asteroseismology determined ages and measured rotation rates, with effective temperatures between 5700-5900 K \citep{vanSaders2016,Creevey2017data}. In addition, we include the stars 18 Sco and $\alpha$ Cen A  with less precisely measured parameters (\cite{vanSaders2016,Metcalfe2022} and  references therein)). Also, we have included a few stars with measured surface magnetic fields and Zeeman Doppler image inferred chromospheric rotation periods from  the Bcool project magnetic survey \citep{Marsden2014}. Note that, compared to the Kepler sample, the Bcool survey does not provide precise photosphere rotational periods; however, it provides more precise measurements for magnetic fields.
We will show  spin evolution solutions for older-than-solar  sun-like stars 1-10 from this data table for both regimes.   Stars 11-15 in the table\ref{table:datapoints} are excluded from our set of solution plots beacuse: stars 11 and 12 are younger than the Sun;  star 13 has a significantly smaller radius, mass and bolometric luminosity than stars 1-10 and is much older than the Sun; 14 is transitioning from the main-sequence to the subgiant phase and 15 is a subgiant. The data points for these disqualified stars are presented only for comparison to the solution curves for stars 1-10.

Fig. \ref{fig:solution} shows the time evolution of the rotation period for individual stars. The top panel shows solutions for regime I, where energy loss due to conduction is dominant and stellar wind energy loss is very low. The bottom panel shows solutions for regime II, where conduction is negligible, and the  X-ray energy losses equal that of the stellar wind.  For all the stars plotted, we chose a coronal temperature $\Tilde{T}_{0,*n}=\frac{1}{2.1}$ for regime I and  $\Tilde{T}_{0,*n}=\frac{1}{1.6}$ for regime II solutions. These exemplify values for which there is a steady Parker wind solution with $r_A>r_s$ for all stars, and for which the values fit within the range of equilibrium temperatures for the solar minimum and maximum \citep{Blackman2016,Johnstone2015}.
   
Overall 
choosing a different value for $\Tilde{T}_{0,*n}$ for both regimes does change the respective slopes of the solutions, but the ranges chosen are consistent with bounds on observed stellar data \cite{Johnstone2015}. If we knew the present X-ray temperature, this would pin down whether a given star is presently in regime I or regime II, and which solution to use. Instead, we compare the consequences of  time evolution solutions from either regime for a given star. We find that the implications are not  that  sensitive to knowing the X-ray temperature over the bounded range because either regime's solutions ultimately lead to our same main conclusions.


Both panels of Fig. \ref{fig:solution} also show the modified Skumanich law \citep{Mamajek2014} $P = t^{0.55}$ and a standard rotational evolution model \citep{vanSaders2013,vanSaders2016}. Regime I curves for stars 1-4 from table \ref{table:datapoints} have
slightly decreasing slopes with age (implying a decreasing  rate of period increase) compared to regime II curves for those stars, whose slopes slightly increase with age (implying an increasing  rate of period increase). This can be identified first by comparing the purple curves in each of the two plots for which the difference is most dramatic. The empirical Skumanich law has decreasing slope with age, akin to Regime I solutions for stars 1-4,  and  captures the data trend quite well.  The rotational evolution model used by \cite{vanSaders2016} has increasing slopes with age, more akin to Regime II. 

Unlike the Skumanich law and \cite{vanSaders2016} curves,  our solutions comprise  an envelope of curves, each passing through a specific star. This envelope is consistent with the observed period-age relation data. In Fig. \ref{fig:solution} blue shaded region corresponds to an envelope of solutions for stars with a more precisely measured rotation period and age. It shows that even without including stars from Bcool project this blue-shaded envelope covers the region with the most stars.  We include the subgiant star data points on the plot (14 an 15 form Table. \ref{table:datapoints}) we do not show their evolution solutions because we are focusing on main sequence stars only and whether the main sequence stars themselves exhibit a spindown transition.
\cite{vanSaders2016} does include the subgiant points in their data fitting, and this strongly affects the shape of their shaded area, which  rises at late times. 

\begin{figure}
\centering
\subfloat{\includegraphics[width=\columnwidth]{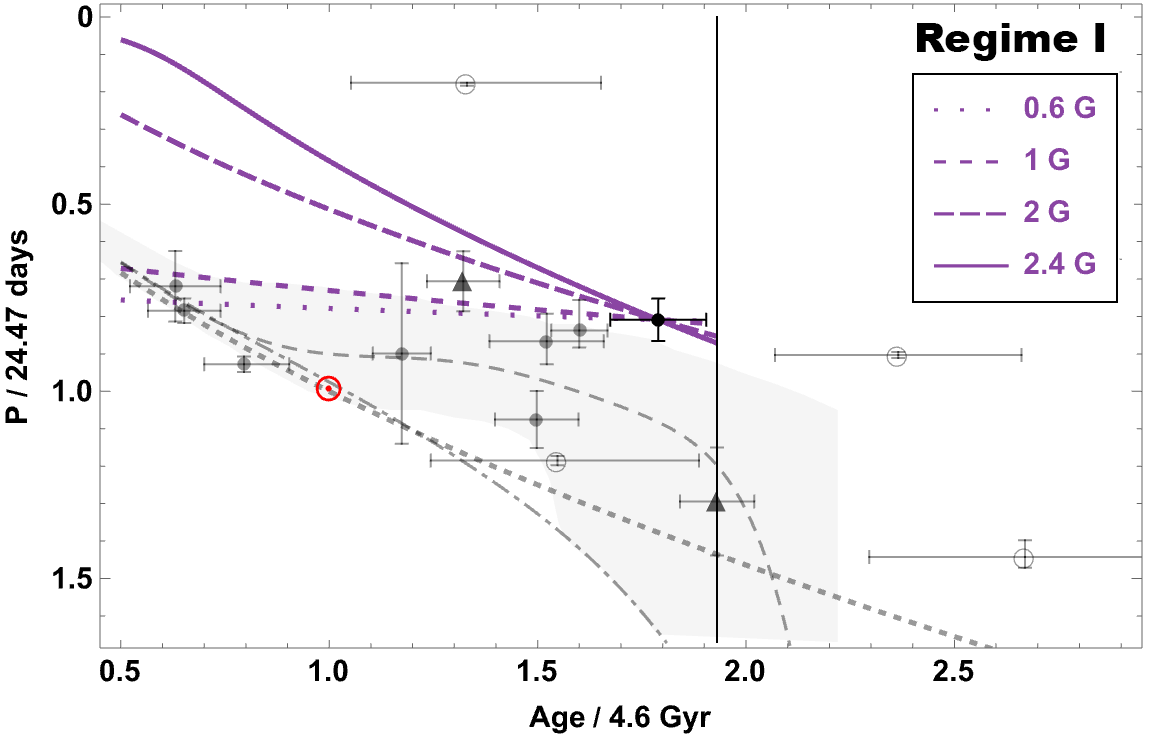}}  \\
\subfloat{\includegraphics[width=\columnwidth]{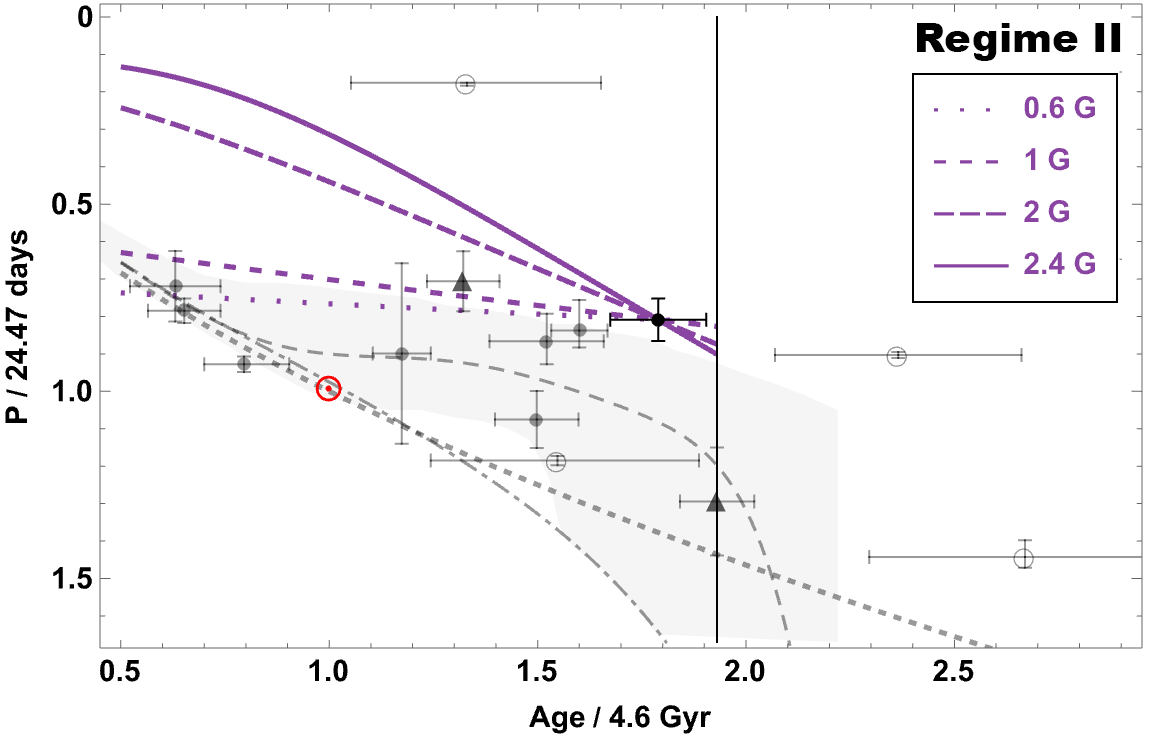}}
 \caption{Panels a and b represent the solutions for the time evolution of the rotational period (purple)  for one specific star (star 2 from the Table \ref{table:datapoints}) to demonstrate the sensitivity of our solutions to magnetic field strength. The top panel corresponds to  regime I and the bottom to regime II .These plots show a significant spread  for different magnetic field strength normalization values for both regimes I and II. Values used for the magnetic field from bottom to top are $B_{p} = 0.6$ G, $1$ G, $2$ G, $2.4$ G, respectively. Data points, black curves (dashed, dotted, dot-dashed) and shaded area have  the same meaning as in Fig \ref{fig:solution}. The vertical line represents the cutoff before the subgiant phase for the stars 1-7 in  Table \ref{table:datapoints}.}
 \label{fig:periodmagfieldreg12}
\end{figure}

\begin{figure}
\centering
\subfloat{\includegraphics[width=\columnwidth]{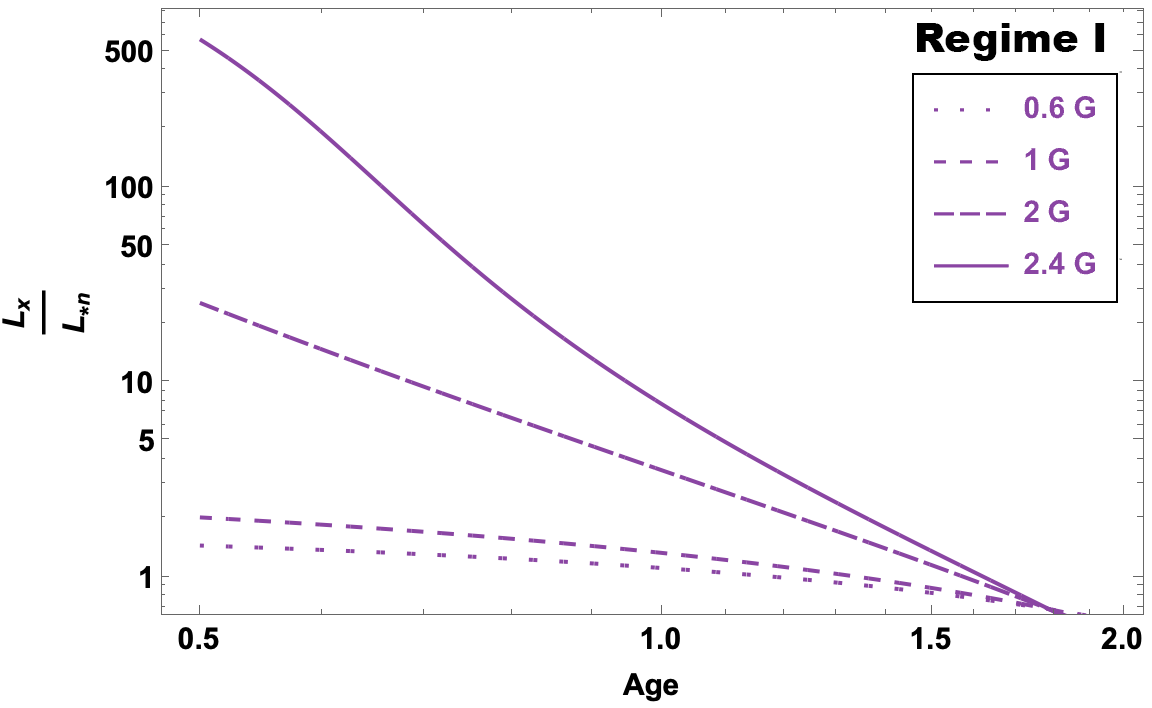}}  \\
\subfloat{\includegraphics[width=\columnwidth]{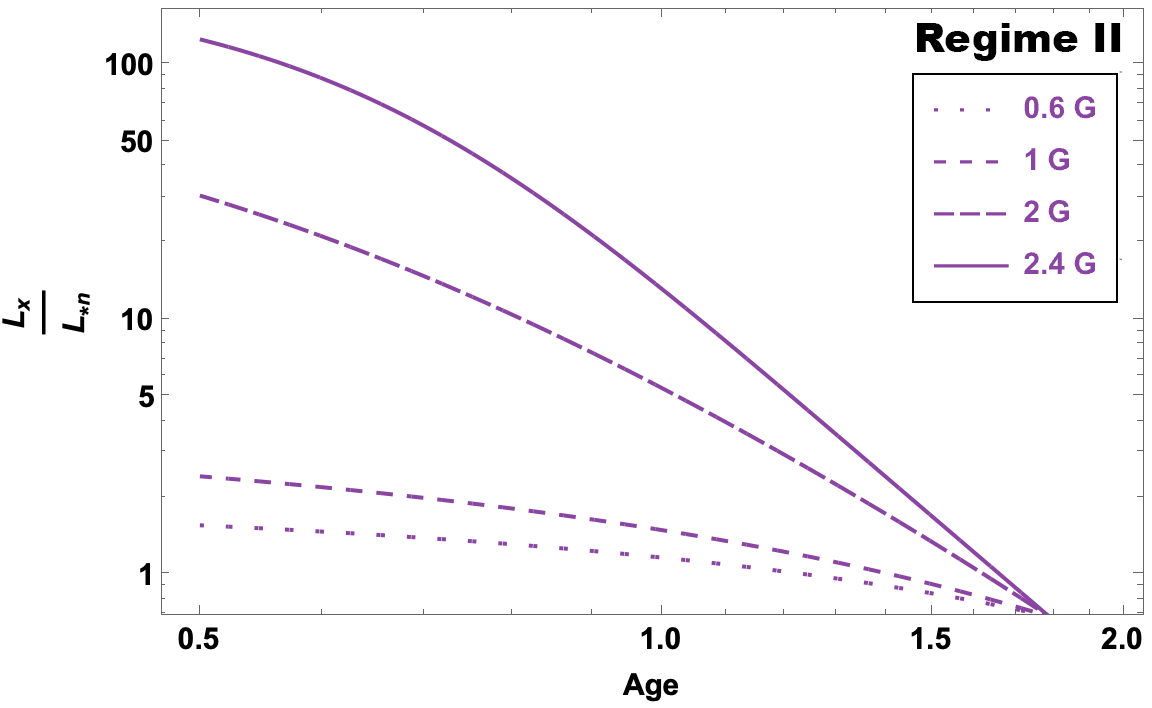}}
 \caption{ The two panels show  solutions for $l_x$ versus time for different magnetic field strengths for star 2 form Table \ref{table:datapoints}. The top panel corresponds to regime I and the bottom to regime II. This spread in the luminosities further demonstrates the sensitivity of our solutions to surface magnetic field strengths. Here we used the same magnetic field values and line styles  as in Fig.\ref{fig:periodmagfieldreg12}.}
 \label{fig:luminomagfieldreg12}
\end{figure}

Observations  do not provide accurate Rossby numbers for stars 2-7 or magnetic fields for stars 2-4. Since  these stars are similar to the Sun in other respects,  for lack of a better option, we simply assume that these quantities  are  comparable to solar values.
Since the magnetic field is the agent of energy transport into the corona, our solutions are quite sensitive to magnetic field strength. To exemplify this
we  present solutions for different magnetic field strengths in Fig. \ref{fig:periodmagfieldreg12} for a star without a  measured magnetic field. The top panel shows solutions for regime I and the bottom panel for  regime II using magnetic field values $B_{p} = 0.6$ G, $1$ G, $2$ G, $2.4$ G. In both regimes we  see the conspicuous difference between solution curves for lower and higher magnetic fields.
Fig. \ref{fig:luminomagfieldreg12}   demonstrates  the  influence of magnetic field strength on $l_x$. 

Generally,   Figures \ref{fig:periodmagfieldreg12} and  \ref{fig:luminomagfieldreg12} show that the broad spread of solutions for the range of magnetic fields considered  makes it difficult to predict the exact evolution path for each star. This  further highlights the imprecision of any prediction for the population that would arise by using one single-line curve. The  theoretical prediction for the population is an  envelope of curves.


\subsection{Physical role of thermal conduction in Regimes I and II}

As mentioned above, we assume that dynamo-produced fields source the coronal energy, which in turn  has three main processes for energy loss: stellar wind, thermal conduction and X-ray radiation. The first two  increase with increasing  temperature, while X-ray radiation decreases. This leads to an equilibrium with a minimum total coronal flux  \citep{Hearn1975}.
For regime I, thermal conduction and  X-ray luminosity dominate the energy loss 
leaving little contribution from the stellar wind. Here conduction removes hot gas available for the wind and the  wind mass-loss rate correspondingly drops exponentially with decreasing gas temperature. This, in turn, reduces the rate of angular momentum loss. 
In regime II, conduction is sub-dominant and wind loss and X-ray radiation dominate the  coronal energy loss.



The difference in increasing and decreasing slopes for stars 1-4 between the two regimes of Figure \ref{fig:solution} that was discussed in the previous subsection  
is caused by the relative influence of thermal conduction, which   is more important at low temperatures where it determines the relation between luminosity and mass loss, and in turn, the coupled evolution of X-ray luminosity, magnetic field strength, and spin.
 
In the spin evolution model used by \cite{vanSaders2016}, the scaling between luminosity and mass loss is the same as in our regime II, equation (\ref{eq:lxm2}), although for different reasons. 
  But there are  key differences  in the approaches and predictions between  our spin evolution model and theirs as we now explain. First, we solve a coupled set of dynamical equations for magnetic field, wind loss, X-ray luminosity and rotation period to produce our solution curves over the entire range of the solution.  In the time ranges that apply to Figure \ref{fig:solution}, our period evolution envelope predicts something close to Skumanich, both for regime I and II in the plots.  For the early time range of this plot, the \cite{vanSaders2016} magnetic breaking formula  also predicts something close to Skumanich.  
  
  However, for the time range beyond $t \sim 0.8 $,  \cite{vanSaders2016} impose by hand an empirically motivated  prescription that the angular momentum is constant and no magnetic breaking occurs.  Instead, we just continue to solve the evolution equations for this regime without making any assumptions of a specific transition.   Their constant angular momentum solution leads to a constant period evolution when the momentum of inertia of the star does not change but for the subgiant phase  it leads to an exponential period increase as the stellar radius expands.  As we are only focused on the main sequence phase, we do not  solve or compare our solutions to their model in this regime. 


\subsection{Influence of  feedback of rotation on  magnetic field evolution}

In regime I,  the relationship between luminosity and mass loss is very different from regime II.   
Regime I overall shows better agreement with the data, although  our envelope of solutions using either regime I or regime II  can describe the observed period-age relation without requiring a change of a dynamo mode. 
 
 Despite the  slight differences in slopes of curves in Regimes I and II previously discussed,  the fact that the solution curves for regime I versus regime II in
 Fig. \ref{fig:solution} do not dramatically differ can be explained by considering the feedback between  rotation and the magnetic field. 
For low mass loss (regime I) the change in the angular momentum, and in turn, the magnetic field is  insignificant, while for regime II stars are losing angular momentum faster, thereby reducing the magnetic field more than in regime I. Because of the dynamical coupling between  the magnetic field and stellar rotation,  reducing the magnetic field also reduces the spin-down rate, resulting in a similar rotation period evolution to that of regime  I.\footnote{Remember that these stars are in the unsaturated regime, where magnetic field and X-ray luminosity do depend on spin.}



\section{Conclusion}\label{sec:conclusion}

To study the time evolution of the stellar rotation period and  the period-age relationship for G and F-type main sequence stars we have employed and generalized a minimalist holistic time-dependent model for spin-down, X-ray luminosity, magnetic field, and mass loss \citep{Blackman2016}. The model combines an isothermal Parker wind \citep{Parker1958}, dynamo saturation model \citep{Blackman2015}, a coronal equilibrium condition \citep{Hearn1975}, and assumes that angular momentum is lost primarily from  the equatorial plane \citep{Weber1967}. 

From a sample of older-than-solar stars chosen  for having precise measurements of period and age, we solved these evolution equations such that each 
star  is a fixed point on a unique solution curve. We argued that the envelope of these curves is a more appropriate indicator of theoretical predictions than a  single line fit  through the Sun or any chosen star to represent the entire population. 

We produce separate such envelopes for cases in which thermal conduction is respectively less or more important and
 both cases, despite slight differences in the curve slopes, are consistent with the data.
Overall,  our  results  
suggest that a dynamo transition from dipole dominated to higher  multipole dominated is not unambiguously required to reduce the rate of spin down, as there is not a clear contradiction between theory and observation for the envelope of solutions without such a transition
when the theory depends on a Parker-type wind solution.

We explored the sensitivity of our solutions to  stellar properties  that we may not know  for individual stars, such as the coronal base X-ray temperature and magnetic field strength. Because the   Parker-type wind solution  is integral to the model,
we are forced to an exponential sensitivity on the coronal base X-ray temperature.
This limits the precision  of any theoretical  or model prediction expressed as a single line  intended to capture the evolution of the stellar population. The prediction should instead be expressed as an envelope of curves.  
Said another way, the sample of observed data does not have enough sufficiently identical stars  to make an ensemble average prediction of high precision.  This connects to the broader need to more commonly express  limitations in precision of theory  field theories applied to astrophysical systems \citep{Zhou+2018}. 

Since it is not possible to  obtain more than 1 data point for individual stars over their spin-down evolution lifetimes, more observations to better nail down evidence for or against a spin-down transition are desired. More data on  individual more closely "identical" stars at different times in their spin-down evolution would be desirable. In addition, at the population level,  period-mass plots for older clusters than have  presently been measured would be valuable. Observations from the Kepler K2 mission have  shown  that by the time  clusters reach an age of 950Myr,  period-mass relations appear to converge to a relatively tight 1 to 1 dependence \cite{GodoyRivera2021}. Similar results were obtained for 2.7 Gyr-old open cluster Ruprecht 147 \citep{Gruner2020}, who found that stars lie in period-mass-age plane with possible evidence for a mass dependence  requiring additional mass-dependent physics parameter variation (perhaps e.g. relating to our $q$ below  Eqn. \ref{eq:omegadot}
deviating from unity),  in modeling  spin-down. 
If similar  data could be obtained for much older clusters and the tight relations were to show strong kinks or bifurcate into more than one branch within the mass range of solar like stars  $0.5<M/M_{\odot}<1.5$, a subset of what we have considered here, this would suggest that the population of  solar-like stars that we are focusing on would show  population-level evidence for a  transition.  


\section{Data Availability}
All the data used in the paper is either created theoretically from equations herein, or given  in Table \ref{table:datapoints}.

\section{Acknowledgments}
KK acknowledges support from a Horton Graduate Fellowship from the Laboratory for Laser Energetics.  We acknowledge support from  the Department of Energy  
grants DE-SC0020432 and DE-SC0020434,
and National Science Foundation grants 
AST-1813298 and PHY-2020249. EB acknowledges the Isaac Newton Institute for Mathematical Sciences, Cambridge, for support and hospitality during the programme "Frontiers in dynamo theory: from the Earth to the stars"where work on this paper was undertaken. This work was supported by EPSRC grant no EP/R014604/1. JEO is supported by a Royal Society University Research Fellowship. This work was supported by the European Research Council (ERC) under the European Union’s Horizon 2020 research and innovation programme (Grant agreement No. 853022, PEVAP). For the purpose of open access, the authors have applied a Creative Commons Attribution (CC-BY) licence to any Author Accepted Manuscript version arising.

\bibliographystyle{mnras}
\bibliography{spinKBO}

\bsp	
\label{lastpage}
\end{document}